\documentclass[preprint2]{aastex}

\newcommand{\e}{\epsilon}
\newcommand{\ep}{\epsilon^\prime}

\newcommand{\epk}{\epsilon_{pk}}
\newcommand{\tp}{t^\prime}

\slugcomment{ApJ, in press, v. 614, 10 Oct., 2004 }

\begin{document}

\title{Curvature Effects in Gamma Ray Burst Colliding Shells}

\author{Charles D. Dermer}
\affil{E. O. Hulburt Center for Space Research, Code 7653,\\
Naval Research Laboratory, Washington, DC 20375-5352}
\email{dermer@gamma.nrl.navy.mil}

\begin{abstract}

An elementary kinematic model for emission produced by relativistic
spherical colliding shells is studied. The case of a uniform
blast-wave shell with jet opening angle $\theta_j \gg 1/\Gamma$ is
considered, where $\Gamma$ is the Lorentz factor of the emitting
shell.  The shell, with comoving width $\Delta r^\prime$, is assumed
to be illuminated for a comoving time $\Delta t^\prime$ and to radiate
a broken power-law $\nu L_\nu$ spectrum peaking at comoving photon
energy $\e_{pk,0}^{\prime}$.  Synthetic GRB pulses are calculated, and
the relation between energy flux and internal comoving energy density
is quantified.  Curvature effects dictate that the measured $\nu
F_\nu$ flux at the measured peak photon energy $\e_{pk}$ is
proportional to $\e^3_{pk}$ in the declining phase of a GRB
pulse. Possible reasons for discrepancy with observations are
discussed, including adiabatic and radiative cooling processes that
extend the decay timescale, a nonuniform jet, or the formation of
pulses by external shock processes. A prediction of a correlation
between prompt emission properties and times of the optical afterglow
beaming breaks is made for a cooling model, which can be tested with
Swift.

\end{abstract}


\section{Introduction}

In the collapsar scenario for GRBs, pulses in GRB light curves are
thought to be produced by collisions between relativistic shells
ejected from a central engine (see \citet{zm04} for a recent
review). The interception of a more slowly moving shell by a second
shell that is ejected at a later time, but with faster speed and
larger Lorentz factor, produces a shock that dissipates internal
energy to energize the particles that emit the GRB radiation. This
scenario is widely considered to explain pulses in GRB light curves
\citep{kps97,dm98}. Studies of pulses are important to decide if 
GRB sources require engines that are long-lasting or impulsive
\citep{dm03}, with important implications for the nature of the central
engine, which is often argued to be a newly formed black hole powered
by the accretion of a massive dense torus.

Here we construct an elementary kinematic model for colliding shells,
assumed spherical and uniform within jet opening angle $\theta_j$.
This is the sort of jet that \citet{fra01} discuss regarding
the standard energy reservoir result, where jet opening angles are
inferred from the time of achromatic spectral breaks in optical
afterglow light curves.

We also perform this study in order to quantify the curvature
constraint of a spherically emitting shell traveling with bulk Lorentz
factor $\Gamma$, which implies that the shell radius
\begin{equation}
r \approx 2\Gamma^2 c t_{var}/(1+z)\;
\label{r}
\end{equation}
in order to produce variability on timescale $t_{var}$
\citep{rl79,fmn96}.  This study also quantifies the rate at which
flux decays at a given energy due to curvature effects, and the range
of validity of the approximate relation
\begin{equation}
\Phi_E \cong c r^2 u_0^\prime \Gamma^2/d_L^2\;
\label{phie}
\end{equation}
between internal comoving energy density $u_0^\prime$ and observed
energy flux $\Phi_E$, where $d_L$ is the luminosity distance (See
Appendix A). The accuracy of this relation is important to quantify
$\gamma\gamma$ opacity constraints \citep{ls01,der04} applied to GRB
pulses as measured with the GRB monitor and Large Area Detector on
GLAST\footnote{glast.gsfc.nasa.gov}, as well as to make estimates of
photomeson production in GRB blast waves
\citep{wb97}.

If curvature effects dominate the late-time emission in GRB pulses,
then a unique relation is found whereby the value of the $\nu F_\nu$
peak flux $f_{\epk}$(in cgs units of ergs cm$^{-2}$ s$^{-1}$) at peak
photon energy $\e_{pk}$ decays in proportion to $\propto\e_{pk}^3$.
This relation is generally not observed in long, smooth GRB
pulses studied by \citet{br01}, who find power-law decays
$f_{\epk}\propto \epk^\zeta$, with $0.6 \lesssim
\zeta \lesssim 3$. Remarkably, values of $\zeta$ for different pulses
within the same GRB are confined to a rather narrow band of values.
The wide range of values of $\zeta$ are found not only in
multi-peaked GRBs, but also in single-peaked GRBs that display smooth
fast-rise, slow-decay light curves \citep{br01,rp02}. The smooth
single peak GRBs could arise from curvature effects
\citep{fmn96}, or to external shocks \citep{dbc99}. For 
GRB pulses that could be produced by spherically symmetric shell
collisions, discrepancy with observations suggest a breakdown of our
assumptions.

In the next section, the kinematic model is presented. Calculations
based on this model are presented in Section 3. In Section 4, we
discuss the possibility that radiative-cooling effects produce the
power-law relation, implying a prediction that can be tested with
Swift\footnote{swift.gsfc.nasa.gov}. Alternately, the uniform
spherical shell assumption could break down, or the basic model of
colliding shells could be in error. The Appendices give derivations of
simple, widely-used approximations related to this study, a derivation
of the curvature relation $f_{\epk} \propto \e_{pk}^3$, as well as an
analytic form for the time-dependent pulse profile, leading to a
simple expression for the light curve of a pulse in the curvature
limit. A brief summary is given in Section 5.

\section{Kinematic Model}

A simple kinematic model for the received flux from the
illumination of a spherically symmetric shell resulting from shell
collisions is studied. A shell with finite width is assumed to be
uniformly illuminated throughout its volume for a fixed duration during
which the shell travels with constant speed from the explosion center.
Light-travel time and Doppler effects are treated without regard to
details of the energization and cooling of the radiating particles.
This approach gives kinematic expectations of curvature effects
in a GRB colliding shell system.

The $\nu F_\nu$ flux measured at dimensionless photon energy $\e =
h\nu/m_ec^2$ and time $t$ is given by 
$$f_\e (t) = {1\over
d_L^2}\;\int_0^{2\pi} d\phi \int_{-1}^1 d\mu \; \int_0^\infty
dr\; r^2 \;\delta^3({\bf r}) $$
\begin{equation}
\times \ep j^\prime(\e^\prime,\mu^\prime,\phi^\prime; {\bf r},t^\prime)\;,
\label{fet1}
\end{equation}
where primes refer to comoving quantities, the integration is over
volume in the stationary (explosion) frame, the Doppler factor
\begin{equation}
\delta\ ={1\over \Gamma(1-\beta\mu)}\;,
\label{delta}
\end{equation}
$\beta = \sqrt{1-1/\Gamma^{2}}$, and $\ep= (1+z)\e/\delta$ (see
\citet{gps99}, noting the correction of a $(1+z)$ factor in the
relation between the emitted and received photon frequencies).
The emissivity $j_*(\e_*,\Omega_* ) = dE_*/dV_*dt_*d\Omega_* d\e_* =
\delta^2 j^\prime(\e^\prime,\Omega^\prime)$, where $\Omega=\Omega_*$
is the directional vector $(\mu,\phi)$, $\mu^\prime =
(\mu-\beta)/(1-\beta\mu)$, and $\phi^\prime = \phi$. We use a notation
where asterisks refer to quantities in the stationary frame (though we
have dropped asterisks for the spatial variables $r$ and $\Omega$),
and unscripted quantities refer to the observer frame.

The blast wave is assumed to emit isotropically in the comoving frame,
which could apply to synchrotron and synchrotron self-Compton
processes with randomly-ordered magnetic fields and electron
pitch-angle distributions, but not to external Compton processes.
Moreover, the observer is assumed to be located along the azimuthal
symmetry axis of the jet, or is viewing a uniform jet with opening
angle $\theta_j
\gg 1/\Gamma$.  Therefore
\begin{equation}
f_\e (t)  = {1\over 2 d_L^2}\;\int_{-1}^1 d\mu 
\;\delta^3\int_0^\infty  dr\; r^2 
\ep j^\prime(\e^\prime; {\bf r},t^\prime)\;,
\label{fet2}
\end{equation}
noting that $\delta({\bf r}) = \delta$ for a uniform jet.  The
emissivity is related to the internal energy density $u_{\e^\prime}({\bf
r},t^\prime)$ through the relation
\begin{equation}
\ep j^\prime(\e^\prime; {\bf r},t^\prime)\cong
 {cu_{\e^\prime}({\bf r},t^\prime)\over \Delta r^\prime}\;,
\label{uep}
\end{equation}
where $\Delta r^\prime = \Gamma \Delta r$ is the proper shell width,
and the mean escape time of photons from the shell volume is
approximated by $\Delta r^\prime /c$.

Further consider a uniform jet with no angular dependence other than that
the emission goes to zero at $\theta \geq \theta_j = \arccos
\mu_j$. The emitting shell is assumed to be illuminated for the
comoving duration $t_0^\prime \leq t^\prime \leq t_0^\prime +\Delta
\tp$. The spectrum is appproximated by a broken power law with peak
$\nu L_\nu$ flux at energy $\ep_{pk}$, given by the expression
$$u_{\e^\prime}({\bf r},t^\prime) = u_0^\prime H(\tp; \tp_0, \tp_0+\Delta
\tp)$$
\begin{equation}
~~~~~~~~~~\times [x^a H(1-x)+x^b H(x-1)]\;,
\label{uep1}
\end{equation}
where $H$ are the Heaviside functions, $a(>0)$ and $b(<0)$ are the
$\nu L_\nu$ indices, and $x = \ep/\ep_{pk,0}=
(1+z)\e/\delta\ep_{pk,0}$. The peak $\nu F_\nu$ comoving photon energy
$\ep_{pk,0}$ is also supposed to be constant throughout the shell. The
total integrated photon energy density for this spectrum is
$u_{tot}^\prime = u_0^\prime (a^{-1}-b^{-1})$.

The observing time $t$ is related to the emitting time measured in the
stationary explosion frame through the relation
\begin{equation}
t_z = {t\over 1+z} = t_*-{r\mu\over c} \; .
\label{time}
\end{equation}
The zero of time $t_* = 0$ corresponds to the moment of shell
ejection, with the location of the inner edge of the shell given by
the relation $r_i(t_*) = \beta c t_*$. The first moment of shell
illumination takes place when the inner edge of the shell is at radius
$r_0 = \beta c t_{*0} = \beta\Gamma c t_0^\prime$ for a shell moving
with constant speed $\Gamma$.

The finite shell width and finite duration of the illumination implies
two constraints on the integrations over $r$ and $\mu$. The
shell-width constraint $\beta c t_* \leq r \leq \beta c t_* +\Delta r$
implies
\begin{equation}
{\beta c t_z\over 1-\beta \mu} 
\leq r \leq {\beta c t_z +\Delta r\over 1-\beta \mu} \;.
\label{r1}
\end{equation}
Due to light-travel time effects of the relativistically moving shell,
 values of $r$ contributing to the signal observed at time $t_z$
 extend over a range $\Delta r/(1-\beta \mu) \sim \Gamma^2
\Delta r$ \citep{ree66,gps99}.

The illumination constraint $\tp_0\leq 
\tp = t_*/\Gamma\leq \tp_0 + \Delta \tp$ implies
\begin{equation}
{1\over \mu}({r_0\over \beta}-ct_z) \leq r 
\leq {1\over \mu}({r_0\over \beta}-ct_z +c \Gamma\Delta \tp) \;.
\label{r2}
\end{equation}
The zero of observer time is when a hypothetical photon ejected at
$t_* =0$ and $r = 0$ from the inner edge of the shell would reach the
observer.  The time at which the signal is first detected by the
observer is therefore given by $t_z^{init} = [(1-\beta)r_0 - \Delta
r]/\beta c\rightarrow {r_0/2\Gamma^2 c} - (\Delta r/c)$, and the
observing time when a photon emitted from the inner edge of the shell
at the first instant of shell illumination reaches the observer is
$t_{z0}=r_0(1-\beta)/\beta c\rightarrow r_0/2\Gamma^2 c$. The final
expressions in these last two relations hold in the limit $\Gamma \gg
1$. Hence 
$$f_\e (t)  = {c u_0^\prime\over 6 d_L^2 
\Delta r^\prime}\;\int_{\mu_j}^1 d\mu  \;\delta^3 (r_u^3 - r_l^3)$$
\begin{equation}
\times [x^a H(1-x)+x^b H(x-1)] \;.
\label{fet3}
\end{equation}
where $r_l = \max[\beta c t_z/ (1-\beta \mu), (r_0/\beta -ct_z)/\mu]$
and $r_u = \min[(\beta c t_z+\Delta r)/ (1-\beta \mu), (r_0/\beta
-ct_z+c\Gamma\Delta \tp)/\mu]$.

\section{Calculations}

We examine the accuracy of the approximate expressions relating $r$
and $t_{var}$, eq.\ (\ref{r}), and energy flux and internal energy
density, eq.\ (\ref{phie}) (see Appendix A).  Let $t_f$ represent a
fiducial variability time scale for the observer. We introduce radius,
time, and width parameters, denoted by $\eta_r$, $\eta_t$, and
$\eta_\Delta$, respectively, to relate $t_f$ to source-frame
quantities. The curvature constraint for blast-wave radius suggests that
we write
\begin{equation}
r = 2\eta_r\Gamma^2 c t_f/(1+z)\;.
\label{etar}
\end{equation}
Because $d\tp = \delta
dt/(1+z)$, we define 
\begin{equation}
\Delta \tp = 2\Gamma \eta_t t_{f}/(1+z)
\label{etat}
\end{equation}
to relate the intrinsic variability time scale $\Delta t^\prime$ in
the comoving frame to $t_f$.  The comoving width of the emitting
region $\Delta r^\prime
\lesssim \Delta \tp/c$, by causality requirements (if it were larger, 
then large amplitude variability would not be possible except for random
statistical fluctuations).  Thus we define
\begin{equation}
\Delta r^\prime = 2\Gamma \eta_\Delta c t_{f}/(1+z)\;,
\label{etadelta}
\end{equation}
with the causal requirement $\eta_\Delta \lesssim \eta_t$. When $\Delta
r^\prime \ll c\Delta \tp$, the duration of the emitting region is not
determined by its causal size scale, but rather by the duration of
emission radiated from a region much smaller than $\Delta \tp/c$.

We solve eq.\ (\ref{fet3}) for the following standard parameters:
 $\Gamma = 300$; $z = 1$ (so that $d_L = 2.02\times 10^{28}$ cm for a
 $\Lambda$CDM cosmology with $\Omega_m = 0.27$, $\Omega_\Lambda =
 0.73$, and Hubble's constant of 72 km s$^{-1}$ Mpc$^{-1}$, as implied
 by the WMAP results; \citet{spe03}); $\ep_{pk,0} =
 (1+z)\e_{pk,0}/2\Gamma$ with $\e_{pk,0} = 1$ (i.e., peak photon
 energy at the beginning of the pulse equal to 511 keV); $u_0^\prime =
 1$ erg cm$^{-3}$, $a = 4/3$; $b = -1/2$; and $t_{f} = 1$ second.

\begin{figure}[t]
\epsscale{1.0}
\vskip-2.0in
\plotone{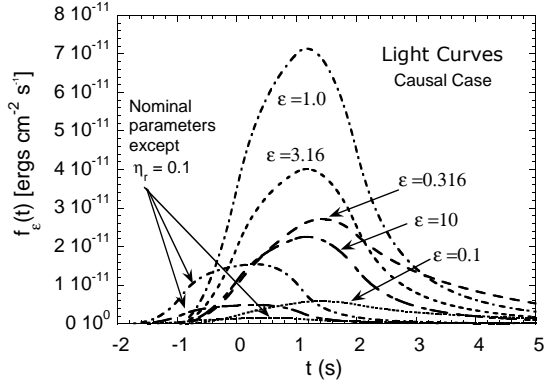}
\caption{Light curves of a causal pulse at different dimensionless
observing energies, for a model with standard parameters (see text)
and $\eta_r=\eta_t = \eta_\Delta =1$.  Also indicated by the arrows
are light curves at $\e = 0.1, 1.0,$ and 10 for a model with $\eta_t =
\eta_\Delta =1$ and $\eta_r = 0.1$.  }
\label{f1}
\end{figure}

Fig.\ 1 shows the appearance of a kinematic pulse with $\eta_r=\eta_t
= \eta_\Delta =1$ at a number of photon energies.  Also shown in Fig.\
1 are kinematic pulses formed when $\eta_t = \eta_\Delta =1$, and
$\eta_r = 0.1$.  Note the characteristic rounded, weakly asymmetrical
(on a linear scale) light curve shapes that are formed when $\Delta
r^\prime \cong c\Delta \tp$. Time delays from different parts of the
width of the emitting shell are important to determine the pulse
shape in this case. The smaller emitting volume when the shell is
energized at $0.1 r_0$ rather than at $r_0$ produces a pulse with a
fluence smaller by a factor $\int_{\eta_\Delta r_0}^{\eta_\Delta
r_0+c\Gamma \Delta t^\prime}dr r^2/\int_{r_0}^{r_0 + c\Gamma \Delta
\tp} dr r^2 \approx $ $\int_{\eta_\Delta r_0}^{\eta_\Delta r_0+r_0}dr
r^2/\int_{r_0}^{2r_0} dr r^2 \approx (1.1^3 - 0.1^3)/(8-1)
\cong 1/7$ smaller. Indeed, $1/7$ is the asymptotic limit of the
fluence reductions due to the different volumes illuminated for a
flare lasting for the equal proper times, but in one case emerging
from deep within the jet, and in the other case where the illumination
begins at the location $r_0 \cong 2\Gamma^2c t_{f}/(1+z)$.  The
intrinsic duration of the pulse, when combined with the curvature
effects, has produced a pulse with FWHM duration of $\approx 2$
seconds, as compared with the fiducial time scale of 1 second. Thus
the combined width, duration, and (off-axis) curvature effects have
lengthened the basic timescale by a factor of about 2 at $\e \cong
\e_{pk,0}$, with a narrower FWHM duration when $\e \gtrsim \e_{pk,0}$
and a broader FWHM duration when $\e \lesssim \e_{pk,0}$.

\begin{figure}[t]
\epsscale{1.0}
\vskip-2.0in
\plotone{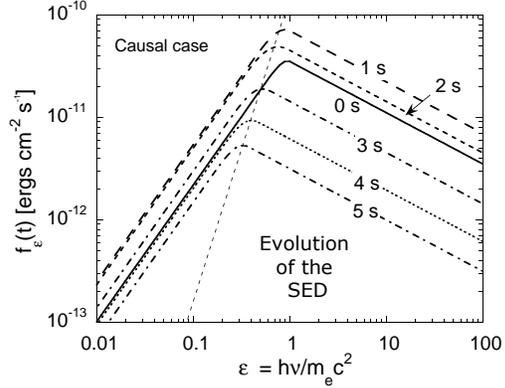}
\caption{Evolution of the spectral energy distribution due to 
curvature effects for a model with standard parameters (see text). In
the declining phase of the pulse, the value of $f_{\e_{pk}} \propto
\epk^3$, as indicated by the dashed line.}
\label{f2}
\end{figure}

Fig.\ 2 shows the evolution of the spectral energy distribution for
this pulse. Notice the rapid decay $\propto \e_{pk}^3$ of the $\nu
F_\nu$ peak flux $f_{\e_{pk}}$ measured at $\epk$ during the decay
portion of the pulse. This behavior is characteristic of all pulses
where curvature effects from off-axis emitting regions dominate the
late-time behavior of the light curve.

Fig.\ 3 shows characteristic light curves when $\Delta r^\prime \ll c
\Delta \tp$, that is, when the shell is very thin compared with the 
size scale associated with the intrinsic pulse duration. These light
curves exhibit pulses that have much sharper peaks than in the general
case of Fig.\ 1, and that are asymmetrical with a distinct trailing
edge of emission. The fluence contained in the pulse is the same as in
the pulse of Fig.\ 1 with $\eta_\Delta = 1$, but the FWHM pulse
duration at $\e = 1$ is $\approx 1$ s, comparable to $t_f$, while the
peak flux is about twice as large, due to the different geometry. The
difference in geometries of a causal and thin-shell pulses introduces
a physical effect required for accurate calculations of scattering or
opacity processes in GRB blast waves.  When $\Delta r^\prime \approx c
\Delta \tp$, then the photon field can be considered to be roughly
isotropic for scattering and opacity calculations.  But when the shell
radiates for a much longer time than the light-crossing time through
the width of the shell, that is, when $\Delta\tp \gg \Delta
r^\prime/c$, the geometry of the outflowing photon flux is much more
anisotropic, giving higher thresholds and lower rates for
$\gamma\gamma$ and photohadronic processes due to the reduction in the
frequency of head-on collisions.

\begin{figure}[t]
\epsscale{1.0}
\vskip-2.0in
\plotone{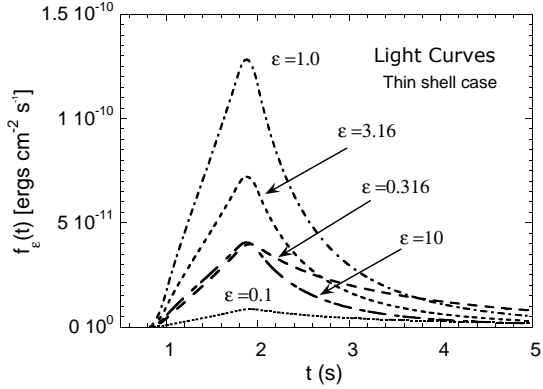}
\caption{Light curves of a thin-shell pulse 
at different dimensionless observing photon energies for a model with
$\eta_r = \eta_t = 1$ and $\eta_\Delta = 0.1$, so
 that the shell width $\Delta r^\prime = 0.1 \Delta\tp/c$.}
\label{f3}
\end{figure}

Fig.\ 4 shows the characteristic light curve shapes formed when
curvature effects dominate the temporal evolution of the light curve
\citep{fmn96}. Here the pulses are very
asymmetrical, with a sharp leading edge. In this calculation, $\eta_r
= 1$ and $\eta_\Delta = \eta_t = 0.1$. The peak $\nu F_\nu$ flux at
$\e = 1$ reaches a value of only $1.4\times 10^{11}$ ergs s$^{-1}$
with a duration of $\approx 0.4$ s. The total energy released is
smaller by a factor of 10 than in Fig.\ 1 with $\eta_{\Delta}$ and
Fig.\ 2 as a result of the shorter intrinsic pulse duration.

The bottom panel in Fig.\ 4 shows, in a
log-log relation, that the flux decays as $t^{-3+b}$ at $\e >
\epk$. When $\e < \epk$, the flux decays as $t^{-3+a}$ at early times,
breaking to a $t^{-3+b}$ behavior at late times due to curvature
effects.

\begin{figure}[t]
\epsscale{1.0}
\plotone{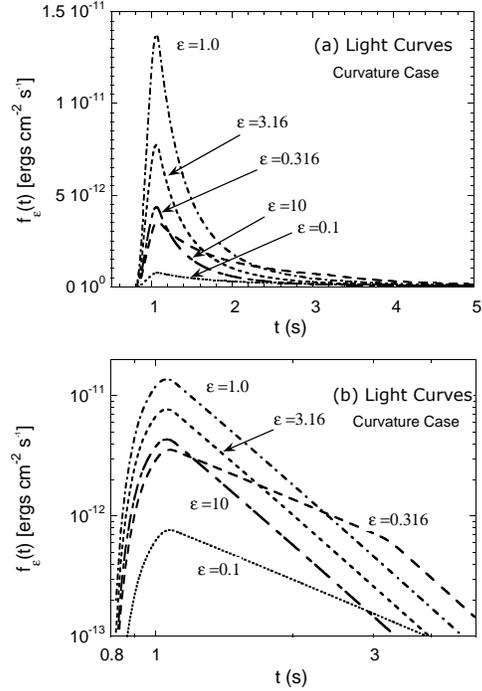}
\caption{ Light curves of a curvature pulse at different 
dimensionless observing photon energies for a model with 
$\eta_r = 1$, but with $\eta_\Delta = \eta_t = 0.1$.  }
\label{f4}
\end{figure}

The spectral and temporal behavior of the 
curvature pulse can be derived in the $\delta$-function
approximation (not to be confused with the Doppler factor $\delta$).
Letting $\ep j^\prime (\ep;{\bf r},\tp) \propto \e^{\prime a}
\delta(r^\prime-r_0)\delta(\tp -\tp_0)= \e^{\prime a}
\delta(r-r_0)\delta[t -(1+z)\tp_0/\delta]$ in eq.\ (\ref{fet2}), 
using the invariance of the 4-volume, then
$$f_\e(t) \propto {r^2_0\over
2d_L^2}\;\e^a\;\int d(1-\beta\mu) \delta^{3-a}
\delta[(1-\beta\mu) - {t_z\over \Gamma \tp_0}]$$
\begin{equation}
\propto \e^a ({\beta c t_z \over r_0})^{-3+a}\propto 
 \e^a({\beta c t_z \over r_0})^{-2-\alpha}\;,
\label{fet4}
\end{equation}
where $\tp_0 = r_0/\beta\Gamma c$, and $\alpha$ is the energy index.
(This result corrects the expressions given by \citet{fmn96} and
\citet{rp02}, where the delta-function pulse in time,
$\delta(\tp-\tp_0)$, is not transformed between the comoving
and observer frames.) The dependence in eq.\ (\ref{fet4}) is derived
more carefully in Appendix B, and analytic forms for the pulse
profile, including a simple functional form for the pulse profile in
the curvature limit, are derived in Appendix C.

\section{Discussion}

The estimate $L \cong 4\pi d_L^2 \Phi_E \cong 4\pi r_0^2 c u_0^\prime
\Gamma^2$, where the received flux is intensified by two powers of
$\Gamma$ for the relativistic time contraction and photon energy
enhancement in a blast wave geometry, is generally used to relate
bolometric energy flux and internal energy density (Appendix A).  More
remarkably, the allowed radius of the radiating spherical shell is
$\approx 2 \Gamma^2 $ times larger than inferred through causality
arguments applied to the measured variability time scale. This effect
greatly dilutes the comoving photon density compared with a stationary
emitting region, and essentially explains the unusual properties of
GRBs. In total, we see that
\begin{equation}
\Phi_{E} \cong {cr_0^2 u_0^\prime\Gamma^2\over d_L^2}=
{4c^3 t_{var}^2\over (1+z)^2 d_L^2}\;u_0^\prime \Gamma^6\;
\propto u_0^\prime \Gamma^6.
\label{phie1}
\end{equation}
For the nominal parameters used in the figures, $\Phi_E \cong
4.8\times 10^{-11} t_{var}^2 ({\rm s})$ $u_0^\prime \Gamma_{300}^6$
ergs cm$^{-2}$ s$^{-1}$.

The most rigorous limits on $\gamma\gamma$ attenuation are obtained by
determining the {\it minimum value} of the product $u_0^\prime \Delta
\tp$ that can produce a pulse with a
measured peak flux $f_{\e_{pk}}$ and full-width half-maximum duration
$t_{1/2}$ for a given value of $\Gamma$. It is the product $u_0^\prime
\Delta \tp$ that enters into the $\gamma\gamma$ attenuation (and
photomeson) calculations. If the shell is found to be optically thick
at some photon energy for a given value of $\Gamma$ even in this case,
then $\Gamma$ must be larger if photons with the corresponding
energies are detected.

Inspection of the various cases shows that the pulse formed in the
curvature limit produces the brightest measured flux and shortest
duration for a given value of the product $u_0^\prime \Delta
\tp$. This is because the measured duration is due entirely to 
curvature effects, and the radiated energy is compressed into the
shortest duration and brightest pulse in this limit. From the results
of Appendices C and D, this implies that the expression
\begin{equation}
u_0^\prime \Delta \tp \cong {[2^{1/(3-a)}-1](1+z) d_L^2 f_{\e_{pk}} 
\over 8 c^3 \Gamma^5 t_{1/2}}
\label{u0cdt}
\end{equation}
gives the smallest possible values for $u_0^\prime c \Delta \tp$, and
this expression will therefore yield the most reliable minimum Lorentz
factors for $\gamma\gamma$ attenuation calculations derived from {\it
GLAST} or ground-based air Cherenkov telescope observations.  The
corresponding expression for the comoving photon spectral energy
density is thus
\begin{equation}
u^\prime_{\e^\prime} \cong
{0.26(1+z) d_L^2 f_{\e_{pk}}
\over 8 c^3 \Gamma^5 t_{1/2} \Delta \tp} [x^a H(1-x)+x^b H(x-1)]\;,
\label{ueprime}
\end{equation}
where $t_{1/2}$ is determined at photon energies near the peak
of the $\nu F_\nu$ spectrum. Eq.\ (\ref{ueprime}) is  a factor of 
3 smaller than the expression used for a comoving spherical blob
with $\delta \rightarrow \Gamma$ and $t_{var} \rightarrow t_{1/2}$
(see Eq.\ [2] in \citet{der04}).

Three generic types of pulses have been identified for the simple
kinematic pulse, namely the curvature case where $r_0 \gg c\Gamma\Delta
\tp$, the causal case where $r_0\approx\Gamma\Delta 
r^\prime \approx c\Gamma\Delta
\tp$, and the thin shell case where $\Delta r^\prime \ll c\Delta \tp$.
In all three types of kinematic pulses, curvature effects dominate the
formation of the spectrum at late time $t \gg (1+z) \Delta
\tp/2\Gamma$, so that $f_{\e_{pk}} \propto \epk^3$ if curvature
effects dominate pulse formation at late times. 

The curvature relationship can be derived from a simple scaling
argument by noting that the differential stationary-frame shell volume
which contributes to the received flux, given by $dV =2\pi r_0^2
\Delta r d\mu$, remains constant with time. This is because the
relation between reception time $t$ and $\mu$ for a shell that is
instantaneously illuminated at comoving time $\tp_0$ is $t =
(1+z)\Gamma\tp_0(1-\beta\mu)$, so that $d\mu
\propto dt$. The $\nu F_\nu$ flux $f_\e =\delta^4 L^\prime/4\pi d_L^2 =
\delta^4 V^\prime \ep j(\ep )/4\pi d_L^2 =
\delta^3 V \ep j(\ep )/4\pi d_L^2$, where $L^\prime$ is the 
comoving luminosity of the emitting volume that contributes to the
flux at time $t$. For an emission spectrum that is flat, that is, $
\ep j(\ep) \propto \e^{\prime 0}$, $f_{\e_{pk}} \propto \epk^3$
because $\epk \propto \delta$ for a uniform shell. 

Analysis of BATSE GRB light curves (Borgonovo and Ryde 2001) shows
that the peak fluxes of a GRB pulse generally follow a relation
whereby
\begin{equation}
f_{\e_{pk}}\propto \epk^\zeta\;.
\label{fepk}
\end{equation}
Values of $\zeta$ for different GRBs vary over a wide range from
$\approx 0.6$ to $3$, with values of $\zeta$ roughly constant for
pulses within the same GRB or in a GRB consisting of a single
smooth pulse.  In most GRBs, therefore, curvature effects do not make
a large contribution to the decay phase of a GRB light curve.

An interesting question is the source of the difference of
observations from our kinematic model pulses. One possibility is that
the jet has angular structure, and varies with directional energy
release and baryon-loading on angles $\theta \approx $ few$\times
\Gamma^{-1}$. The angle-dependent speeds in such a system 
would produce a deformed colliding shocked fluid shell where the
spherical symmetry assumption fails, as therefore would the uniform
jet model. If this is the case, then GRB prompt emission data can in
principle be analyzed to reveal shell structure and to determine
whether this behavior is consistent with a universal jet structure
(\citet{zha04}; see
\citet{fra03} for review).

Rather than treat these geometrical effects here, we consider instead
whether radiation effects could form a power-law relation between
$f_{\e_{pk}}$ and $ \epk$. The most naive system considers a fixed
volume of shocked fluid within which the peak of the $\nu F_\nu$
spectrum is made by a large population of quasi-monoenergetic
electrons that radiates most of the power through the synchrotron
process in a mean magnetic field of strength $B$. If these electrons
mainly have comoving Lorentz factors $\gamma$, then their luminous
power $\propto B^2\gamma^2$. Because the peak of the $\nu F_\nu$
spectrum is $\propto B \gamma^2$, and assuming $B$ is constant, then
$ f_{\e_{pk}}\propto \epk$ or $ \zeta_{syn} = 1$ for this simple
synchrotron model with constant magnetic field. This model can
therefore only apply in rare cases.
 
A better treatment must consider the evolution of $\gamma$ due to
synchrotron and adiabatic losses in the expanding shell.
The equation of electron energy evolution is given by
\begin{equation}
-{d\gamma\over d\tp} = {1\over V^\prime_{sh}} {dV^\prime_{sh}\over
d\tp}\; {\gamma\over 3} + {\sigma_{\rm T}B^2(\tp )\over 6\pi
m_ec}\;\gamma^2\;,
\label{dgammadt}
\end{equation}
where the comoving shell volume changes with time according to
$V^\prime_{sh} \propto t^{\prime 3m}$, with $m = 0$ corresponding to
no expansion, and $m = 1$ corresponding to 3-dimensional expansion.

The magnetic field will also change as a result of the expansion of
the shell volume. In the flux freezing limit where the magnetic field
is randomly oriented, $BR^2 \propto const$, implying $B \propto
V_{sh}^{\prime 2/3}\propto t^{\prime -2 m}$. The well-ordered magnetic
field required to explain the polarization observation of the GRB
021206 observed with RHESSI \citep{cb03} suggests that there is not
an efficient mixing and randomization of the magnetic field
directions. For simplicity, we therefore write $B \propto t^{\prime -2
v m}$, where $v = 1$ gives the flux-freezing limit.

Eq.\ (\ref{dgammadt}) becomes
\begin{equation}
-{d\gamma\over d\tau} = m\; {\gamma\over \tau} + \nu_0\tau ^{-4vm}\gamma^2\;,
\label{dgammadt1}
\end{equation}
where $\tau \geq 1$ is a dimensionless time variable, and $\nu_0$ is a
dimensionless synchrotron energy loss rate. Eq.\ (\ref{dgammadt1}) is
analytic, but it is sufficient to consider two limiting cases of
dominant adiabatic losses or dominant synchrotron losses at late
times. In the case of dominant adiabatic losses we have (dropping the
primes) $\gamma\propto t^{ -m}$, $B\propto t^{ -2 v m}$, and $\e_{pk}
\propto t^{-2m (1+v)}$, so that $ f_{\e_{pk}}/ \epk \propto B \propto 
\e_{pk}^{v/(v+1)}$. Thus $\zeta_{adi} = 1 + [v/(v+1)]$. Even for a wide 
range of values of $v$, $1 \lesssim \zeta_{adi} \lesssim 2$, and 
$\zeta_{adi}$ is independent of the geometry factor $m$.

If synchrotron losses dominate the cooling, $-d\gamma/dt \propto B^2
\gamma^2$.  The dependence $B\propto t^{ -2 v m}$ therefore implies
$\gamma \propto t^{4vm-1}$, so that $\e_{pk} \propto B\gamma^2 \propto
t^{6vm-2}$. Hence $f_{\e_{pk}}/\e_{pk} \propto B \propto
\e_{pk}^{vm/(1-3vm)}$, so that $\zeta_{syn}= 1+ [vm/(1-3vm)]$. Except when
$v \ll 1$, $\zeta_{syn} \approx 1$ when $m=0$ and $\zeta_{syn} \approx
0.5$ -- 0.67 when $m = 1$. In this simple model, therefore, 
values of $\zeta$ between $1/2$ and $2/3$ are only possible for 
3-dimensional expansion.

Three-dimensional expansion is more likely to occur for narrowly
collimated blast waves than for blast waves with large opening angles,
and the narrowly collimated blast waves would have ``beaming breaks"
in the optical afterglow light curves at earlier times. If our
conjecture that the $f_{\e_{pk}}$/$\e_{pk}$ relationships are due to
synchrotron and adiabatic effects in GRB blast waves with different
opening angles, then those blast waves with $\zeta < 1 $ should be
correlated with earlier beaming break times. Because this effect is
only seen when synchrotron losses dominate the cooling, GRBs with
$\zeta < 1$ should also display cooling spectra with photon indices
$\approx 3/2$ below $\e_{pk}$. 

\citet{br01} find several GRBs and many pulses in the BATSE sample 
with statistically significant values of $\zeta $ less than
unity. These GRBs however preceded the afterglow era. For those GRBs
which have measured beaming breaks (see Table 1 in \citet{bfk03}),
only GRB 990123 has sufficiently bright BATSE data to provide a data
point for such a correlation. GRB 990123 has not yet been analyzed to
give $\zeta$, while analysis of Beppo-SAX data is in progress (F.\
Ryde, private communication, 2004).

Such a model for the $f_{\e_{pk}}$/$\e_{pk}$ relationship
would explain why $\zeta$ is approximately constant for different
pulses within a GRB, provided that the opening angle of the GRB jet
remains the same throughout the period of activity of the GRB engine.

The adiabatic/synchrotron model would not, however, explain pulses with $2
\lesssim \zeta \lesssim 3$. There are many such pulses in the
\citet{br01} sample, though generally with large error bars. If analysis
of Beppo-SAX or Swift data reveal such GRBs, then another explanation
is required.  One possibility is that GRB pulses are due to the
interactions of a single impulsive blast wave with inhomogeneities in
the surrounding medium. This version of the external shock model for
the prompt phase can be much more efficient than an internal shell
model \citep{dm99,dm03}, and permits quantitative studies of the
statistics of BATSE GRBs (\citet{bd00}; see \citet{zm04} for a review
of the internal/external controversy).

Predictions for the $f_{\e_{pk}}/\e_{pk}$ relationship in an
external shock model \citep{dcb99} can be derived by adapting the
equations for blast wave deceleration in a uniform medium with $\Gamma
= \Gamma_0/[1+(x/x_d)^g]$, where $\Gamma_0$ is the initial Lorentz
factor, $x_d$ is the deceleration distance and $g$ is the radiative
index ($g = 3/2$ and 3 for an adiabatic and fully radiative blast
wave, respectively). In the deceleration phase, $x \propto
t^{1/(2g+1)}$ and therefore $\Gamma \propto t^{-g/(2g+1)}$. In this
model, $\e_{pk} \propto \Gamma B \gamma_{pk}^2$ and $f_{\e_{pk}}
\propto \Gamma^2 B^2 \gamma_{pk}^2$, where $\gamma_{pk} \propto
\Gamma^4\propto t^{-4g/(2g+1)}$ in the slow-cooling regime, 
and $\gamma_{pk} \propto (x\Gamma)^{-1}\propto t^{-2/(2g+1)}$ in the
fast-cooling regime. Thus $f_{\e_{pk}}$/$\e_{pk}
\propto B\Gamma$.

In the slow-cooling regime, $\e_{pk} \propto \Gamma^4\propto
t^{-4g/(2g+1)}$, and $f_{\e_{pk}}\propto\e_{pk}^{3/2}$. In the
fast-cooling regime, $\e_{pk} \propto t^{-2/(2g+1)}$ and
$f_{\e_{pk}}\propto\e_{pk}^{1+g}$. In the slow-cooling and
fast-cooling regimes, therefore, values of $\zeta_{sc} = 3/2$ and
$\zeta_{fc} = 1+g$, respectively, are predicted. Provided that the
surrounding medium is uniform (which can be inferred from afterglow
modeling, though at a larger distance scale), the slow-cooling result
implies a definite value of $\zeta_{sc}=3/2$ for fast-rise, smooth
decay light curves when spectral analysis demonstrates that the GRB
evolves in the slow-cooling regime.  For GRBs in fast-cooling regime,
this estimate implies $5/2 < \zeta_{fc} < 4$, and in these cases
cooling spectra should be apparent. Further work will be needed to
extend the results to radial density gradients of the circumburst
medium, and to verify that these relations hold for deceleration in
small density inhomogeneities that form GRB pulses in the external
shock model.

\section{Summary}

A simple kinematic model for GRB colliding shells has been constructed
that provides a framework for analyzing radiative processes in a
simplified geometry of a thin or thick shell traveling at relativistic
speeds.  The relationship between observed flux and comoving photon
energy density for a given value of $\Gamma$ has been studied, showing
that the curvature limit yields the smallest value of the product
$u_0^\prime \Delta \tp$. This result can then be used to deduce
conservative lower limits on bulk Lorentz factors derived from the
condition of $\gamma\gamma$ transparency.

The kinematic model predicts the curvature relationship $f_{\e_{pk}}
\propto \e_{pk}^\zeta$, with $\zeta = 3$, at late times in GRB
pulses. Equivalently, curvature effects imply that $f_\e (t)\propto
t^{-3+a}$ and $\epk \propto t^{-1}$. BATSE data for GRB pulses do not
display the curvature relationship in most cases \citep{br01},
suggesting that the physics of pulse formation is dominated by other
effects. A simple model for joint evolution of $f_{\e_{pk}}$ and
$\e_{pk}$ that takes into account adiabatic and synchrotron losses
implies that $1/2 < \zeta < 2$, and that $\zeta \approx 0.5$ only when
the shell undergoes three dimensional expansion and the electrons
which produce the emission near $\e_{pk}$ are rapidly cooling through
synchrotron losses. Spectral analysis of Swift data, and correlations
of $\zeta$ with times of the beaming breaks in optical afterglow light
curves, can test this prediction. Such a correlation would validate an
adiabatic/synchrotron model for GRB prompt radiation, relate
properties of the prompt phase with the afterglow, and provide key
insights into the properties of GRB jets.

Another possibility is that the $f_{\e_{pk}} \propto \e_{pk}^\zeta$
relationship is formed by external shock processes, and a simple derivation
of $\zeta$ was given for blast-wave deceleration in a uniform surrounding
medium. Analysis of prompt data has the potential to test the external
shock model, though complications regarding density gradients
and inhomogeneities in the circumburst medium must be considered in 
more detail.

A final possibility is that jet structure produces the measured
relationship between $f_{\e_{pk}}$ and $\e_{pk}$. The validity of a
universal jet model will be tested by determining whether observed
values of $\zeta$ can derive from the proposed angle dependence. In
the meantime, comparing the predictions of the adiabatic/synchrotron
and external shock models with time-resolved spectroscopy of GRB
pulses and afterglows has the potential to rule out or validate these
models.

\vskip0.2in
\noindent 
I thank Markus B\"ottcher, Felix Ryde, and the anonymous referee
for valuable comments. This work is supported by the Office of Naval
Research and NASA {\it GLAST} Science Investigation grant DPR
S-13756G.

\appendix

\section{Relations and Estimates}

First we relate the total energy in photons between the comoving and
stationary frames.  The differential number of photons
$N_*(\e_*,\Omega_*)$ per unit energy and solid angle transforms as
$N_*(\e_*,\Omega_*) = \delta N^\prime (\ep, \Omega^\prime)$, as is
easily seen by calculating the Jacobian of the transformation, or by
noting that $\e^{-1}dN/d\e d\Omega \equiv \e^{-1} N(\e,\Omega)$ is an
invariant. For an isotropic, monochromatic photon spectrum in the
comoving frame, $N^\prime (\ep,
\Omega^\prime)= N_0 \delta(\ep - \ep_0)/4\pi$, and the total photon
energy in the comoving frame is just $E^\prime_0 = N_0 \ep_0$ (in
units of the electron rest mass). The differential photon spectrum in
the stationary frame is therefore $N_*(\e_*,\Omega_*) = \delta N_0
\delta(\e_* /\delta -\ep_0)/4\pi$ $ = \delta^2 N_0
\delta(\e_*-\delta\ep_0)/4\pi$, so that 

\begin{equation}
E_* = \oint d\Omega_* \int_0^\infty d\e_* \;\e_* \;N_*(\e_*,\Omega_*) =
{N_o \ep_0\over 2}\int_{-1}^1 d\mu\;\delta^3=\Gamma E^\prime\;.
\label{E*}
\end{equation}
This result is obvious by noting the symmetry of the transformation
equation $\e_* = \Gamma \ep(1+\beta \mu^\prime)$ with respect to
$\mu^\prime$.

Because $\Phi_E = L/4\pi d_L^2$, by definition of the luminosity
distance $d_L$, the fluence $\varphi = \Phi_E\langle t \rangle = L_*
\langle t \rangle/4\pi d_L^2 = L_* \langle t_* \rangle (1+z) /4\pi
d_L^2$, where $\langle t \rangle$ and $\langle t_*\rangle$ are times
of reception and emission in the observer and stationary frame,
respectively. Using eq.\ (\ref{E*}) gives
\begin{equation}
\varphi = {\Gamma E^\prime\over 4\pi d_L^2} (1+z)\;.
\label{varphi}
\end{equation}

A simple estimate relating comoving energy density $u^\prime_0$ with
energy flux $\Phi_E$ is obtained by noting that the stationary frame
luminosity of a blast wave is given by $L_*= dE_*/dt_* = \Gamma^2
L^\prime $, where $L^\prime = dE^\prime /d\tp \cong u_0^\prime 4\pi
r^2 \Delta r^\prime/(\Delta r^\prime/c)$.  Thus $\Phi_E \cong
cr^2 u_0^\prime\Gamma^2/d_L^2$, giving eq.\ (\ref{phie}).  If the
variability is produced by curvature effects according to eq.\
(\ref{r}), then
\begin{equation}
\Phi_E = {4 c^3 u_0^\prime \Gamma^6 t_{var}^2\over (1+z)^2 d_L^2}\;.
\label{phie2}
\end{equation}
Note that the same basic dependence, though
with $\Gamma$ replaced by $\delta$, is derived
for a (comoving) spherical blob geometry. In this case,
 $\Phi_E \cong \delta^4 L^\prime/4\pi d_L^2$, and 
$L^\prime \cong 4\pi r_b^{\prime 2} c u_0^\prime/3$, with 
blob radius $r_b^\prime = c \delta t_{var}/(1+z)$.

\section{Analytic Derivation of $f_{\e_{pk}}(t)$ vs.\ $\e_{pk}$ Relation}

Starting with eq.\ (\ref{fet2}), we approximate
\begin{equation}
\ep j^\prime(\ep;{\bf r},\tp )= K\e^{\prime a} \delta(r^\prime -r_0)
\delta(\tp - \tp_0)H(\ep; \ep_l,\ep_u)\;.
\label{epjp}
\end{equation}
Normalizing to the comoving energy $E_p^\prime$ of a pulse implies
that
\begin{equation}
K = {a E_p^\prime \over 2\pi (1-\mu_j) r_0^2 
(\e^{\prime a}_u - \e^{\prime a}_l)}\;.
\label{K}
\end{equation}

The integrals can now be performed. First note the subtlety that
 $|dr/dr^\prime| = \delta$, whereas $\Delta r^\prime = \Gamma \Delta
 r$ in equation (\ref{uep}). Imposing the limits over $r$ in equation
 (\ref{r1}) recovers the $\delta$ factor in the numerical integration
 of equation (\ref{fet3}) performed in Section 2.  Furthermore noting
 that $\tp_0 = r_0/\beta
\Gamma c$, and defining $\e_z = (1+z)\e$, we obtain
\begin{equation}
f_\e(t) = {K c r_o \e_z^a\over 2 d_L^2 }\;
\bigl( {\beta \Gamma c t_z\over r_0}\bigr )^{-3+a}\times
H[{\beta c t_z \over r_0};\max(1-\beta,{\ep_l\over \Gamma\e_z}),
\min(1-\beta\mu_j,{\ep_u\over \Gamma \e_z})]\propto \e_z^a t_z^{-3+a}\;.
\label{fet30}
\end{equation}
The final proportionality holds provided that $t_z$ is in the range
satisfying the Heaviside function. When $a = 0$, corresponding to
emission at the peak of the $\nu F_\nu$ spectrum,
$f_{\e_{pk}}(t)\propto t_z^{-3}\propto \e_{pk}^3$.  This follows
because $\e_{pk} \propto t_z^{-1}$, as is apparent by inspecting the
limits in the Heaviside function ($\beta c t_z/r_0 \propto
\ep_u/\Gamma\e_z$, so that $\e_{pk} \propto \ep_{pk}/t_z$).

The validity of eq.\ (\ref{fet30}) can be checked by deriving the fluence
$\varphi = \int_0^\infty d\e \; \int_{-\infty}^\infty dt \; f_\e(t)/\e$, 
using normalization (\ref{K}), from which eq.\ (\ref{varphi}) is recovered.

\section{Relationship between $f_\e(t)$ and  
$u_0^\prime$ in the Curvature Limit }

We now derive an approximate analytic expression for a radiation pulse
in the curvature limit. Substituting expressions (\ref{uep}) and
(\ref{uep1}) for the comoving spectral energy density into eq.\
(\ref{fet2}), the $r$-integral can be approximately solved by letting
$\int dr r^2[\dots]/\Delta r^\prime\rightarrow r_0^2 |dr/dr^\prime|
[\dots ]\cong \delta r_0^2 [\dots ]$. In the limit $\Gamma
\gg 1$, one obtains
\begin{equation}
f_\e(t) = {4 c u_0^\prime r_0^2 \Gamma^2\over d_L^2}\;
[{(\e/\e_{pk,0})^a\over 3-a}Q_a
+ {(\e/\e_{pk,0})^b\over 3-b}Q_b]\;,
\label{fetQ}
\end{equation}
where
\begin{equation}
Q_a = [\max(1,{u\over 1+\eta})]^{-3+a} -
 [\min (4\Gamma^2,u,{\e_{pk,0}\over \e})]^{-3+a}\;,
\label{Qa}
\end{equation}
\begin{equation}
Q_b = [\max(1,{u\over 1+\eta},{\e_{pk,0}\over \e})]^{-3+b} - 
[\min (4\Gamma^2,u)]^{-3+b}\;,
\label{Qb}
\end{equation}
$u \equiv 2\Gamma^2 c t_z/r_0$, 
$\eta \equiv\Gamma c \Delta \tp/r_0 = \eta_t/\eta_r$, and
we set $\mu_j = -1$ (otherwise the terms 
$4\Gamma^2$ are replaced with $ 2\Gamma^2(1-\beta\mu_j)$ in
eqs.\ (\ref{Qa}) and (\ref{Qb}) above). The $\nu F_\nu$
peak energy observed at the start of the pulse is denoted by
$\e_{pk,0} =2\Gamma \ep_{pk,0}/(1+z)$.

By examining the limits in eq.\ (\ref{Qa}), one finds that
\begin{equation}
Q_a = \cases{
1 - u^{-3+a}\;,\; & $1 \leq u \leq {\e_{pk,0}\over \e} \leq 1+\eta$ ,\cr
1 - ({\e_{pk,0}\over \e})^{-3+a}\;,\;& $1 \leq {\e_{pk,0}\over \e}
 \leq u \leq  1+\eta $ ,\cr 
({u\over 1+\eta})^{-3+a}- ({\e_{pk,0}\over\e})^{-3+a}\;, & $
1\leq {u\over 1+\eta} \leq  {\e_{pk,0}\over \e}\leq u$ ,\cr
({u\over 1+\eta})^{-3+a}- u^{-3+a}\;,& $ 1+\eta 
\leq u \leq  {\e_{pk,0}\over \e} $ ,\cr }
\label{Qaapprox}
\end{equation}
with related expressions for $Q_b$. In the limit $\eta \ll 1$,
corresponding to the curvature limit where variability arises
principally from curvature effects, the fourth relation
in eq.\ (\ref{Qaapprox}) applies, giving
\begin{equation}
f_\e\cong {4 c u_0^\prime r_0^2 \Gamma^2\over
d_L^2}\;\eta [({\e \over \e_{pk,0}})^a u^{-3+a} H({\e_{pk,0}\over u} - \e )
+ ({\e \over \e_{pk,0}})^b u^{-3+b} H(\e - {\e_{pk,0}\over u})]\;.
\label{curvaapprox}
\end{equation}
This expression applies equally to the late-time asymptote 
$t \gg (1+z)\Delta\tp/2\Gamma$. 
At the peak of the $\nu F_\nu$ spectrum, $a = 0$, and
\begin{equation}
f_{\e_{pk}} = {4 c u_0^\prime r_0^2 \Gamma^2\over d_L^2}\;\eta 
\bigr({2\Gamma^2 c t_z\over r_0}\bigl)^{-3}\;,
\label{fep8}
\end{equation}
recovering the dependence derived in eq.\ (\ref{fet30}).  Note that
because $r_0\propto \eta_r$, $f_\e\propto \eta_r\eta_t$.  Eq.\
(\ref{curvaapprox}) relates $f_\e(t)$ and $u_0^\prime$ in the
curvature limit.

\section{Searching for Curvature Effects in GRB Pulses}

If the GRB spectral flux is described by a power-law spectrum with
$\nu F_\nu$ index $ a$, then curvature effects would produce the
behavior
\begin{equation}
f_\e(t)\propto t_z^{-3+ a }\,.
\label{fe}
\end{equation} 
The FWHM width of the curvature spectrum in such a regime is given by
$t_{1/2}= [2^{1/(3-a)}-1]t_{pk}$, where $t_{pk} = (1+z)r_0/2\Gamma^2
c$. Hence the expression $r_0 = 2\Gamma^2 c t_{FWHM}/\{
[2^{1/(3-a)}-1](1+z)\}$ relates the blast wave radius $r_0$ to
$\Gamma$, given the observables $z$ and $t_{FWHM}$---provided that the
pulse shape is determined by curvature effects.  Curvature effects
also dictate that $f_{\e_{pk}}\propto \e_{pk}^3$.  By examining the
variation of intensity as a function of $\e_{pk}$ for two GRBs,
\citet{sf01} searched for the signature of shell curvature using an
expression analagous to eq.\ (\ref{fe}), though without success. In
this case, $f_{\e_{pk}}\propto \e^{3-\langle a+b\rangle}$, where
$\langle a+b\rangle$ is the mean index of the photon flux within the
interval containing $\e_{pk}$ used to measure $f_{\e_{pk}}$ (to first
order, $\langle a+b\rangle = 0$).

Eq.\ (\ref{fep8}) can be used to derive an expression relating
$\e_{pk}$ to the photon fluence $\varphi$ in the curvature limit.
One simply obtains
\begin{equation}
{ \e_{pk}\over \e_{pk,0}} = \sqrt{1-{\varphi\over \varphi_{tot}}}\;,
\label{epkvarphi}
\end{equation}
where $\e_{pk,0}$ refers to the value of $\e_{pk}$ at the
beginning of the pulse, and $\varphi_{tot}$ refers to the total
fluence. This expression represents an alternative analytic form to
the relation $\e_{pk}/\e_{pk,0}=\exp{(-\varphi/ \varphi_{tot})}$
proposed by \citet{lk96}, and also derives from eq.\ (\ref{curvaapprox}),
provided that $a$ and $b$ are independent of time.

When expression (\ref{epkvarphi}) deviates from observational data, as
will often be the case since the approximations leading to the
curvature pulse are rarely expected to be realized in GRB colliding
shells, then curvature effects can still be sought by numerically
evaluating eqs.\ (\ref{fet3}) or (\ref{fetQ}) to obtain more general
$\e_{pk}$-$\varphi$ relations. These equations can also be used to fit
pulse profiles directly. Such an approach would place the
phenomenological treatments of \citet{krl03} and \citet{ryd04} on a
physical basis, and can be extended to treat realistic electron
injection and loss processes.  Such results can then be compared with
predictions of the external shock models for the prompt phase.

\bibliography{your bib file}


\end{document}